\documentclass[aps,pra,twocolumn,showpacs]{revtex4-1}
\usepackage{amsbsy,latexsym,amsmath,amsthm,amsfonts,mathrsfs,bbold}
\usepackage{amsfonts}
\usepackage{amssymb}
\usepackage{hyperref}
\usepackage{epsfig,graphics,graphicx}
\usepackage{color}
\usepackage{enumerate}

\newcommand{\two}[2]{\begin{array}{c}\\[-1.5em]\scriptstyle #1\\[-.3em] \scriptstyle #2\end{array}}
\newcommand{\vecc}[1]{\mbox{{\boldmath $#1$}}}

\begin{document}

\title{A generalized wave-particle duality relation for finite groups}
\author{Emilio Bagan$^{1,3}$, John Calsamiglia$^{3}$, J\'anos A. Bergou$^{1,2}$, and Mark Hillery$^{1,2}$}
\affiliation{$^{1}$Department of Physics and Astronomy, Hunter College of the City University of New York, 695 Park Avenue, New York, NY 10065 USA \\ 
$^{2}$Graduate Center of the City University of New York, 365 Fifth Avenue, New York, NY 10016 \\ 
$^{3}$F\'{i}sica Te\`{o}rica: Informaci\'{o} i Fen\`{o}mens Qu\`antics, Departament de F\'{\i}sica, Universitat Aut\`{o}noma de Barcelona, 08193 Bellaterra (Barcelona), Spain}

\begin{abstract}
Wave-particle duality relations express the fact that knowledge about the path a particle took suppresses information about its wave-like properties, in particular, its ability to generate an interference pattern.  Recently, duality relations in which the wave-like properties are quantified by using measures of quantum coherence have been proposed.  Quantum coherence can be generalized to a property called group asymmetry.  Here we derive a generalized duality relation involving group asymmetry, which is closely related to the success probability of discriminating between the actions of the elements of a group.   The second quantity in the duality relation, the one generalizing which-path information, is related to information about the irreducible representations that make up the group representation. 
\end{abstract}

\maketitle

\section{Introduction}
Resource theories, in particular the resource theory of coherence, have been an area of considerable recent activity.  In a resource theory, one has a set of free states, which do not possess the resource, and free operations that do not create the resource.  In addition, there is a measure of the extent to which a state that is not a free state does possess the resource.  The first such theory was that of entanglement.  In that case, the free states are the separable states.  In the case of coherence, one specifies a basis, and the free states are those that are diagonal in that basis \cite{baumgratz}.

The resource theory of coherence is an example of a broader class of resource theories that are characterized by asymmetry under a group of transformations \cite{gour,marvian}.  One starts with a group, $G$, and a unitary representation of the group, $U(g)$ for $g\in G$ acting on a Hilbert space~$\mathcal{H}$.  States, $\rho$, that are invariant under the action of the group, i.e. $U(g)\rho\, U^{\dagger}(g) = \rho$ for all $g\in G$, constitute the free states, and the free operations are those that satisfy $\mathcal{E}[U(g)\rho\, U^{\dagger}(g)]=U(g)\mathcal{E}(\rho)U^{\dagger}(g)$ for all $g\in G$ and all $\rho$, where $\mathcal{E}$ is a completely positive, trace preserving map.  Maps with this property are called G-covariant \cite{marvian}.  States for which ${\mathcal U}_g(\rho):=U(g)\rho\, U^{\dagger}(g) \neq \rho$ for at least one $g\in G$, are said to possess asymmetry.  The resource theory of coherence results when the group is taken to be a cyclic group.

A useful measure of asymmetry is the robustness of asymmetry \cite{piani1,piani2}.  For a given state $\rho$, it is given by
\begin{equation}
\mathcal{A}_{\mathcal{R}}(\rho )= \min_{\tau\in\mathcal{D}(\mathcal{H})} \left\{ s\geq 0 \left| \frac{\rho+s\tau}{1+s} \in \mathcal{S}\right. \right\} ,
\end{equation}
where $\mathcal{D}(\mathcal{H})$ is the set of density matrices on $\mathcal{H}$ and $\mathcal{S}$ is the set of free states.   It has the following useful property.  If one is trying to discriminate among the states $U(g)\rho\, U^{\dagger}(g)$ for $g\in G$, and each of the states is equally probable, the robustness of asymmetry of $\rho$ is closely related to the optimal minimum-error probability of successfully discriminating among the states, $P_{\rm s}(\rho )$.  In particular, we have that \cite{piani2}
\begin{equation}
P_{\rm s}(\rho ) = \frac{1+ \mathcal{A}_{\mathcal{R}}(\rho ) }{|G|},
\end{equation}
where $|G|$ is the number of elements in $G$.  This relation suggests that in this scenario, i.e., discriminating among the equally probable states $U(g)\rho\, U^{\dagger}(g)$, $P_{\rm s}(\rho )$ itself is a good measure of asymmetry.  It has a clear operational  interpretation.  It tells us how good a state is for discriminating the quantum channels ${\mathcal U}_g$.  In channel discrimination, one sends an input state into a channel, and then discriminates as best one can, the output states~\cite{dariano1,chiribella}.  In general the input states can be in a Hilbert space that is larger than the one the channel acts on, but we will only consider states in the carrier space for the representation~$U(g)$.  If $P_{\rm s}(\rho )$ is small, then the state~$\rho$ is a poor input state to use for channel discrimination, which means that its asymmetry must be small, too. If $P_{\rm s}(\rho )$ is close to one, then it is a good input state and also very asymmetric. It is also the case that $P_{\rm s}(\rho )$ has some additional properties that are desirable for a measure of asymmetry.  It decreases under G-covariant quantum maps and it is convex.  These properties follow from those of the robustness of asymmetry, but, for completeness, short proofs are provided in Appendix~\ref{App A}. 

\begin{figure}[htbp]
\begin{center}
$$
\includegraphics[width=25em]{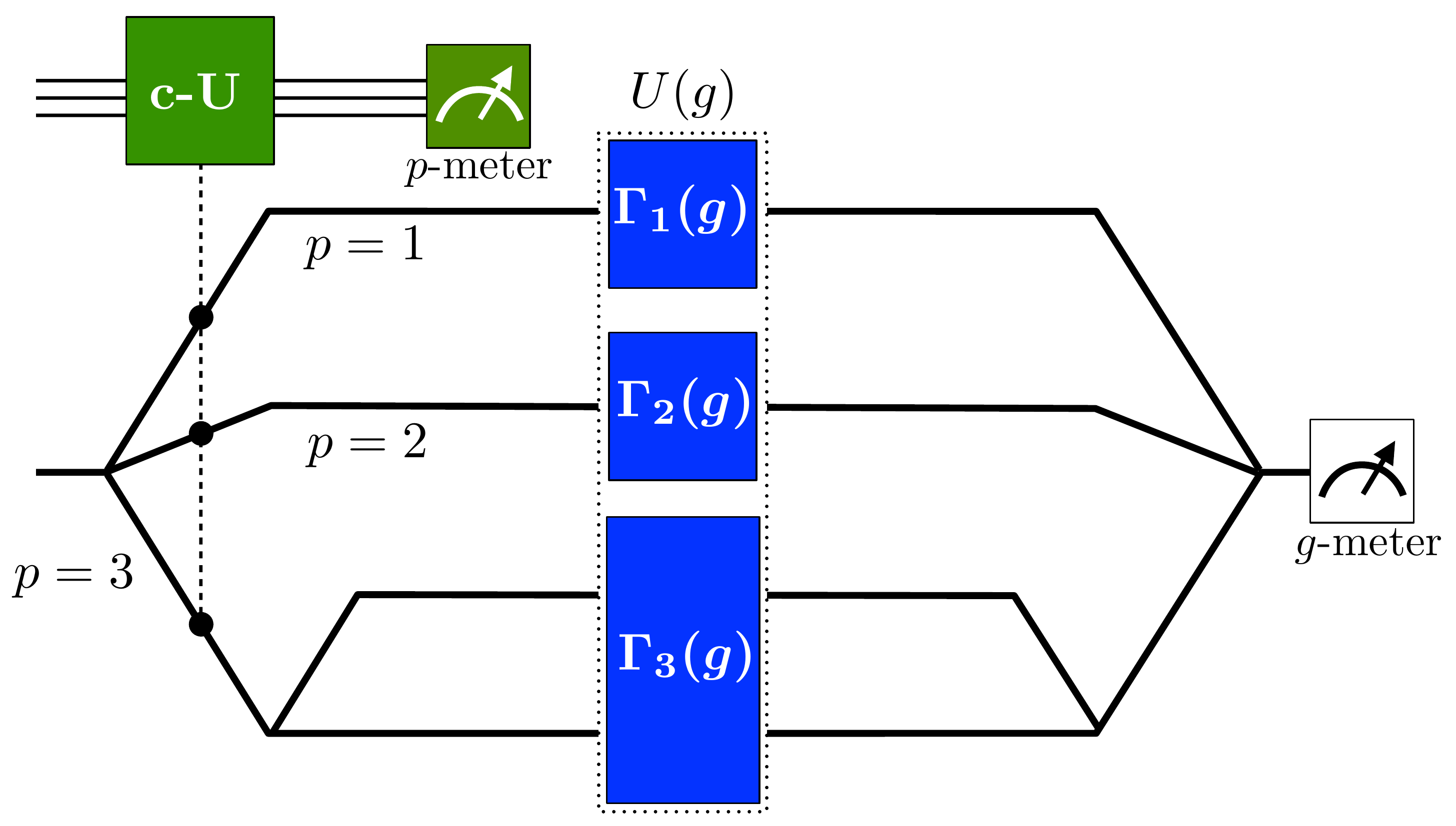}
$$
\caption{Schematic representation of a network for the second example in Sec.~\ref{sec examples}, $U(g)=\oplus_{p=1} ^3\Gamma_p(g)$, on a multiport interferometer. An ancillary system (three lines on top) is used to tag, through a controlled-unitary gate (labeled c-U), the parts of the network on which the irreducible representations $\{\Gamma_p\}_{p=1}^3$~act.
The discrimination of the ancillary states by the optimal measurement (labeled $p$-meter) tells us which part of the network the photon went through with minimum-error success probability~$P_{\rm s}$. The $g$-meter represents a measurement intended for determining which $g\in G$ has been implemented by the blue boxes (e.g., by analyzing the interference pattern or otherwise).
}
\label{f-1}
\end{center}
\end{figure}

In a wave-particle duality experiment, a particle goes through an interferometer, and there are detectors that provide some information about which path the particle took.  There is a tradeoff, expressed by the duality relation, between how much information one has about the path and the visibility of the interference pattern produced by the particle \cite{wootters,greenberger,jaeger,englert,durr,bimonte,englert2,jakob,englert3,angelo}. The higher the visibility, the easier it is to discriminate among different phases imprinted to the particle state by, e.g., phase-shift plates placed in the paths.  In the case considered here, the paths, or rather the orthogonal one-dimensional subspaces that represent~them, are replaced by the invariant subspaces that carry the irreducible representations, and the phases by the channels ${\mathcal U}_g$.  If one tags these subspaces with ancillary states, which can be thought of as detector sates and are not, in general, orthogonal, we find that the probability of discriminating the tagging states places a limit on the probability of discriminating the channels~${\mathcal U}_g$, for $g\in G$.  
The tagging states, then, affect the asymmetry of the input state, 
by affecting the coherence between the different subspaces.  This duality notion for asymmetry could be implemented, e.g.,  by an optical network such as that in Fig.~\ref{f-1}, where detectors tell whether a photon has gone through different parts, the parts corresponding to the subspaces.  The tags are then the states of the detectors.  Our complementarity relation can be viewed as providing a tradeoff between being able to identify which network we have, expressed by $U(g)$ [in the figure $U(g)$ is the direct sum of three irreducible representations $\Gamma_1(g)$,  $\Gamma_2(g)$, $\Gamma_3(g)$, of dimension~1, 1 and 2, respectively], and knowing which part of a network the photon went through.

The paper is organized as follows.  In Section~\ref{sec 2} we present and discuss a formula for $P_{\rm s}(\rho )$ in the case that~$\rho$ is a pure state and there are no repeated irreducible representations.
In Sec.~\ref{sec 3} we derive a duality relation in the simplest case where irreducible subspaces have the same probability, i.e., the particle can be found in each part of the network with equal probability. The general case, including also that of  irreducible representations with multiplicity greater than one and the possible use of entanglement with an idler particle, is left for a separate publication~\cite{us next}.

\section{A simple expression for the success probability}
\label{sec 2}
The discrimination of states generated by the action of a representation of a group acting on a single state, i.e., the states $\{U(g)|\phi\rangle\, |\, g\in G\}$, has been studied by a number of people.  The case of cyclic groups was treated by Ban \emph{et al.}\ and this was extended to abelian groups by Eldar and Forney \cite{ban,eldar}.  The problem for general groups has been studied by Chiribella \emph{et al.} \cite{chiribella1,chiribella2} and Krovi \emph{et al.} \cite{krovi}. We shall make use of a formula for the probability of successfully discriminating among the states $\{U(g)|\phi\rangle\, |\, g\in G\}$ with minimum error that was obtained in~\cite{krovi}. For completeness, we present a proof in Appendix~\ref{App B}.

Suppose that when the representation $U(g)$ is expressed as a direct sum of irreducible representations, each irreducible representation appears at most once.  For any state $|\phi\rangle$, we then have
\begin{equation}
\label{Ps}
P_{\rm s}(|\phi\rangle\langle\phi |) = \left( \sum_{p} \sqrt{\frac{d_{p}}{|G|}} \|\phi_{p}\| \right)^{2} .
\end{equation}
Here the sum is over the irreducible representations that appear in $U(g)$, $d_{p}$ is the dimension of the p$^{\rm th}$ irreducible representation, and $|\phi_{p}\rangle$ is the component of $|\phi\rangle$ in the subspace, ${\mathcal H}_p$, that carries the  p$^{\rm th}$ irreducible representation.  Note that this relation plus the convexity property of~$P_{\rm s}$ (See Appendix~\ref{App A}) can be used to find an upper bound on $P_{\rm s}$ for mixed states.

We can use the above expression to find the best pure state to discriminate the channels $U(g)$ by maximizing the right-hand side.  The Schwarz inequality and the fact that $\sum_{p} \|\phi_{p}\|^{2}=1$ imply that
\begin{equation}
P_{\rm s}(|\phi\rangle\langle\phi |) \leq \frac{1}{|G|} \sum_{p} d_{p} ,
\label{Ps coh}
\end{equation}
and that this bound is achieved when 
\begin{equation}
\|\phi_{p}\| = \left(\frac{d_{p}}{\sum_{p^{\prime}}d_{p^{\prime}} } \right)^{\kern-.3em1/2} .
\end{equation}
To attain the success probability in Eq. \eqref{Ps coh} coherence among the various irreducible subspaces is required. If no such coherence exists, the maximum success probability is given by Eq.~(\ref{D}) below.

\section{A duality relation}
\label{sec 3}
There are duality relations that limit one's ability to both know which path a particle took and to produce an interference pattern with that particle.  More recently, duality relations originating from entropic uncertainty relations~\cite{coles1,coles2} or incorporating coherence measures have been derived~\cite{pati,bagan}.  As we have mentioned and will show in Sec.~\ref{sec examples}, the $l_{1}$ coherence measure is closely related to the optimal success probability of discriminating among states generated by the action of a cyclic group.  This suggests that it should be possible to find a duality relation for more general groups.

Let us consider a representation $U(g)=\bigoplus_{p=1}^{N} \Gamma_{p}(g) $, where each irreducible representation appears at most once, and a pure state of the system as input, given by
\begin{equation}
|\psi\rangle_{\rm S}=\frac{1}{\sqrt{N}}\sum_{p=1}^N |u_p\rangle_{\rm S},
\label{psi & u's}
\end{equation}
where $|u_{p}\rangle_{\rm S}$ is a normalized state in the subspace ${\mathcal H}_p$ corresponding to $\Gamma_{p}$. We use an ancillary system to tag the~$N$ subspaces by applying a controlled-unitary gate to system plus ancilla, the latter having been prepared in an initial state $|\eta_0\rangle_{\rm A}$. If $\openone_{p}$ is the projector onto the invariant subspace ${\mathcal H}_p$, then the gate has the form \mbox{$\sum_{p=1}^N\openone_p\otimes V_p$}, where the unitaries $\{V_p\}_{p=1}^N$ acting on the ancillary system are such that $V_p|\eta_0\rangle_{\rm A}=|\eta_p\rangle_{\rm A}$. The resulting state in ${\mathcal H}={\mathcal H}_{\rm S}\otimes{\mathcal H}_{\rm A}$, ${\mathcal H}_{\rm S}=\oplus_{p=1}^N{\mathcal H}_p$, is
\begin{equation}
|\Psi\rangle= \frac{1}{\sqrt{N}}\sum_{p=1}^{N} |u_{p}\rangle_{\rm S} |\eta_{p}\rangle_{\rm A} .
\label{Psi}
\end{equation}
To simplify the notation, we will drop the indexes ${\rm S}$ (system) and ${\rm A}$ (ancilla) wherever no confusion may arise.  The ancillary states $\{ |\eta_{p}\rangle\}_{p=1}^N$ are normalized but, in general, not orthogonal.  If the channel ${\mathcal U}_g$ is applied, the state becomes
\begin{equation}
|\Psi_g\rangle=\frac{1}{\sqrt N}\sum_{p=1}^N [U(g)|u_p\rangle] |\eta_p\rangle.
\end{equation}
We note that the tagging and the channel application commute, so tagging after the channel application would lead to the same result. Let
\begin{eqnarray}
\rho_g& = & {\rm Tr}_{\rm A}\left(|\Psi_g\rangle\langle\Psi_g |\right) \nonumber \\
& = & \frac{1}{N} \sum_{p,p^{\prime}=1}^{N}\!U(g)|u_{p}\rangle \langle u_{p^{\prime}}|U^{\dagger}(g) \,\langle\eta_{p^{\prime}}|\eta_{p}\rangle \nonumber \\
& = & U(g)\rho_{e}U^{\dagger}(g) ,
\label{rho_g}
\end{eqnarray}
where $\rho_{e}$ corresponds to the identity element of the group. We want to find a relation between our ability to discriminate the states $\{ \rho_{g}\}_{g\in G}$ and our ability to discriminate the states $\{ |\eta_{p}\rangle\}_{p=1}^N$.

With no loss of generality, we can discriminate the states $\{\rho_{g}\}_{g\in G}$ with a covariant POVM $\{\Pi_g\}_{g\in G}$, where $\Pi_{g}=U(g)\Pi_{e}U^{\dagger}(g)$, and $\Pi_{e}$ is the POVM element corresponding to the identity element of the group, $e\in G$.  This implies that our probability of successfully discriminating the channels ${\mathcal U}_g$ with the input state $|\psi\rangle$ is 
\begin{equation}
P_{{\mathcal U}_g}:=\frac{1}{|G|} \sum_{g\in G}{\rm Tr}(\Pi_{g}\rho_{g}) = {\rm Tr}(\Pi_{e}\rho_{e}) .
\end{equation}
%
Now, using that $\Pi_e\ge 0$,
\begin{eqnarray}
\hspace{-1.5em}{\rm Tr}(\Pi_{e}\rho_{e})\! & = &\! \frac{1}{N}\!\!\sum_{p,p^{\prime}=1}^{N}\!\!\langle u_{p^{\prime}}|\Pi_{e}|u_{p}\rangle\langle \eta_{p^{\prime}}|\eta_{p}\rangle \nonumber \\
\!& \leq &\! \frac{1}{N}\!\!\!\sum_{p,p^{\prime}=1}^{N}\!\!\!\!\sqrt{\!\langle u_{p^{\prime}}\!|\Pi_{e}|u_{p^{\prime}}\!\rangle\!\langle u_{p}|\Pi_{e}|u_{p}\rangle} |\langle \eta_{p^{\prime}}|\eta_{p}\rangle| .
\end{eqnarray} 
%
%
From Appendix B, we have that 
\begin{equation}
\langle u_{p}|\Pi_{e}|u_{p}\rangle \leq \frac{d_{p}}{|G|} ,
\end{equation}
so that
\begin{equation}
\label{pi-rho}
P_{{\mathcal U}_g}=
{\rm Tr}(\Pi_{e}\rho_{e}) \leq \frac{1}{N|G|} \sum_{p,p^{\prime}=1}^{N} \sqrt{d_{p}d_{p^{\prime}}}
|\langle \eta_{p^{\prime}}|\eta_{p}\rangle| .
\end{equation}
This inequality can be satisfied as an equality in some circumstances.  Let us choose $\Pi_{e}=|X\rangle\langle X|$, where $|X\rangle = \sum_{p=1}^{N} |X_{p}\rangle$, and $|X_{p}\rangle = e^{i\theta_{p}}\sqrt{d_{p}/|G|}|u_{p}\rangle$ is the component of $|X\rangle$ in the carrier space of $\Gamma_{p}$.  This will satisfy $\sum_{g\in G} \Pi_{g} = \openone_{\rm S}$ (see Appendix~\ref{App B}), and we then have that 
\begin{equation}
{\rm Tr}(\Pi_{e}\rho_{e}) = \frac{1}{N|G|} \sum_{p,p^{\prime}=1}^{N} \sqrt{d_{p}d_{p^{\prime}}}
e^{i(\theta_{p^{\prime}}-\theta_{p})} \langle \eta_{p^{\prime}}|\eta_{p}\rangle  .
\end{equation}
If the $\{ \theta_{p}\}_{p=1}^N$ can be chosen so that 
\begin{equation}
e^{i(\theta_{p^{\prime}}-\theta_{p})} \langle \eta_{p^{\prime}}|\eta_{p}\rangle= | \langle \eta_{p^{\prime}}|\eta_{p}\rangle | ,
\end{equation}
then the inequality in Eq.\ (\ref{pi-rho}) will be satisfied as an equality. One case in which this inequality becomes an equality is the case in which the vectors $\{ |\eta_{p}\rangle\}_{p=1}^N$ are orthonormal.  This implies that
\begin{equation}
D:=P_{{\mathcal U}_g}^{\rm orth}=\frac{1}{N|G|} \sum_{p=1}^{N} d_{p} .
\label{D}
\end{equation}
It follows that in this case $|\Psi\rangle$ is maximally entangled, and thus $\rho_g$ is the least informative state, i.e., it has minimum asymmetry,  among those in Eq.~(\ref{Psi}).  Hence, $D$ is the minimum value of $P_{{\mathcal U}_g}$. We read off from Eq.~(\ref{D}) that $D=1/|G|$ (random guessing) if all irreducible representations are one dimensional (as is the case of coherence). We will come back to this below.  


Now let us move on to the duality relation. In addition to $P_{{\mathcal U}_g}$, the relation involves  the probability that one correctly infers the value of $p$ from the discrimination of the tagging states. In the example of Fig.~\ref{f-1}, this would amount to inferring which part of the network the particle went through.  Tracing out the system, we find that the state of the ancilla is
\begin{equation}
\rho_{\rm A}={\rm Tr}_{\rm S}\left(|\Psi_g\rangle\langle\Psi_g|\right)=\frac{1}{N}\sum_{p=1}^N |\eta_p\rangle\langle\eta_p| ,
\end{equation}
independently of the channel ${\mathcal U}_g$ that has been applied.
Clearly, $\rho_{\rm A}$ corresponds to the specific ensemble where each of the states $\{|\eta_p\rangle\}_{p=1}^N$ is equally likely.  
The optimal probability~of discriminating them with minimum error, $P_{{\mathcal H}_p}$, satisfies~\cite{bagan},  
\begin{equation}
P_{{\mathcal H}_p}-\frac{1}{N} \leq \frac{1}{N^{2}} \sum_{p,p^{\prime}=1}^{N} \sqrt{1-|\langle \eta_{p}|\eta_{p^{\prime}}\rangle |^{2}} .
\label{P_p}
\end{equation}
Let us define the two-component vectors
\begin{equation}
\vecc v_{pp'}\!:=\!\left( \!\frac{1}{N} \sqrt{1\!-\!|\langle \eta_{p}|\eta_{p^{\prime}}\rangle |^{2}},\ \frac{\sqrt{d_{p}d_{p^{\prime}}}}{|G|} |\langle \eta_{p}|\eta_{p^{\prime}}\rangle | \right) ,
\end{equation}
such that
\begin{eqnarray}
\|\vecc v_{pp'}\|^{2} \!& = &\! \frac{1}{N^{2}}\!\! \left[ 1\!+\! \left(\! \frac{N^{2}d_{p}d_{p^{\prime}}}{|G|^{2}}\!-\!1\!\right) |\langle \eta_{p}|\eta_{p^{\prime}}\rangle |^{2}\right] \nonumber \\
\!& \leq &\! \frac{M}{N^{2}} , 
\end{eqnarray}
where $M$ is the maximum over all possible sets $\{|\eta_p\rangle\}_{p=1}^N$ of the term in square brackets. An obvious upper bound for $M$~is
\begin{equation}
M\le \tilde M: =1+ \max_{\two{p,p^{\prime}}{ p\neq p^{\prime}}}\left\{\left( \frac{N^{2}d_{p}d_{p^{\prime}}}{|G|^{2}}-1 \right), 0 \right\}  .
\end{equation}
We then have from Eqs.~(\ref{pi-rho}) and~(\ref{P_p}) that
\begin{eqnarray}
\left(P_{{\mathcal U}_g}-D\right)^{2}  +  \left(P_{{\mathcal H}_p}-\frac{1}{N}\right)^{2} \leq \nonumber \\
 \frac{1}{N^{2}} \sum_{\two{p,p^{\prime} =1}{p\neq p^{\prime}}}^{N} \sum_{\two{q,q^{\prime}=1}{q\neq q^{\prime}}}^{N}  \vecc v_{pp'}\cdot\vecc v_{qq'} .
\end{eqnarray}
Making use of the Schwarz inequality we find 
\begin{equation}
\left(P_{{\mathcal U}_g}-D\right)^{2}  +  \left(P_{{\mathcal H}_p}-\frac{1}{N}\right)^{2} \leq  M\left( 1-\frac{1}{N}\right)^{2} .
\end{equation}
This is the desired duality relation which constitutes the central result of the paper.  It places a limit on our ability to tell which channel, ${\mathcal U}_g$, we have and which invariant subspace ${\mathcal H}_p$ the particle went through. 

\section{Examples} 
\label{sec examples}

\subsection{Cyclic group}
\label{subsec cyclic}
Let us first look at the case in which $G$ is the cyclic group of order $N$.  Let $a$ be the generator of the group, and then the group is $G=\{ a^{n}\}_{\raisebox{.05em}{$\scriptstyle n=0$}}^{N-1}$.  We have that~\mbox{$a^{0}=e$}, the identity element, and~$a^{N}=e$.  The irreducible representations of $G$ are one-dimensional, and there are~$N$ of them.  We shall denote the elements of the~p$^{th}$ irreducible representation by $\Gamma_{p}(a^{n})$.  If the state~$|u_p\rangle$ transforms according to the p$^{th}$ irreducible representation, then
\begin{equation}
\Gamma_{p}(a^{n})|u_p\rangle = e^{2\pi i pn/N}|u_p\rangle .
\end{equation}
Now consider the representation
\begin{equation}
U=\bigoplus_{p=0}^{N-1} \Gamma_{p} .
\end{equation}
This is an $N$-dimensional representation of $G$, and its carrier space is spanned by the orthonormal states~$\{ |u_p\rangle\}_{p=0}^{N-1}$. 

For a vector $|\phi\rangle$ in the carrier space of $U$, we have, from Eq.\ (\ref{Ps}) that
\begin{equation}
P_{\rm s} (|\phi\rangle\langle\phi |)= \frac{1}{N}\left( \sum_{p=0}^{N-1} |\langle u_p|\phi\rangle |\right)^{2} .
\end{equation}
If we now set $\rho=|\phi\rangle\langle\phi |$ and subtract $1/N$, which is just the probability of guessing which state $U(a^{n})|\phi\rangle$ we have, we obtain
\begin{equation}
P_{\rm s}(\rho)-\frac{1}{N} = \sum_{\two{p,q=0}{p\neq q}}^{N-1}| \langle u_p|\rho |u_q\rangle |
= \sum_{\two{p,q=0}{p\neq q}}^{N-1}| \rho_{pq} |.
\end{equation}
This is just the $l_{1}$ measure of coherence defined in \cite{baumgratz} in the basis $\{u_p\}_{p=0}^{N-1}$.  Physically we can interpret the states~$|u_p\rangle$ as corresponding to different paths in an interferometer and the factors of $\exp (2\pi i pn/N)$ as resulting from phase shifters placed in those paths.

\subsection{\boldmath Non-Abelian groups: dihedral group $D_3$ or symmetric group~$S_3$}
\label{subsec non-abelian}

Next, let us look at a non-abelian group.  A simple non-abelian group is the dihedral group $D_{3}$, which consists of rotations and reflections in the plane that leave an equilateral triangle invariant.  It has six elements, $\{ e,r,r^{2}, s, rs, r^{2}s \}$, where $r^{3}=e$ and $s^{2}=e$.  The dihedral group $D_3$ is isomorphic to the symmetric group~$S_3$, i.e., the group of permutations of three elements. The mapping is defined  by $s\mapsto(12)$, $r\mapsto(123)$.
The group has three conjugacy classes $C_{e}=\{ e\}$, $C_{r}=\{ r,r^{2}\}$, and~$C_{s}=\{ s, rs, r^{2}s \}$.  It has three irreducible representations, $\Gamma_p$ for $p=1,2,3$, where $\Gamma_1$ and $\Gamma_2$ are one-dimensional and $\Gamma_3$ is two dimensional.  The character table for the group is given in Table~\ref{t-1} for completeness.

\begin{table}
\centering
\begin{tabular}{|c|c|c|c|} \hline
 & $C_{e}$ & $C_{r}$ & $C_{s}$ \\ \hline $\Gamma_1$ & $1$ & $1$& $1$ \\ \hline $\Gamma_2$ & $1$ & $1$ & $-1$ \\ \hline $\Gamma_3$ & $2$ & $-1$ & $0$ \\ \hline 
\end{tabular}
\caption{\label{t-1}Character table for $D_{3}$.}
\end{table}

The one-dimensional representations are the trivial representation, $\Gamma_1(g)=1$ for all $g\in D_3$, and the so-called sign or alternate representation, defined by $
\Gamma_2(r)=1$ and $\Gamma_2(s)=-1$ for the generators of the group $r$ and $s$. 
For the representation $\Gamma_3$, we can take the matrices,
\begin{equation}
\Gamma_3(r)=\left( \begin{array}{cc} -1/2 & -\sqrt{3}/2 \\ \sqrt{3}/2 & -1/2 \end{array} \right),\ \Gamma_3(s)=\left( \begin{array}{cc} 1 & 0 \\ 0 & -1 \end{array}\right) ,
\end{equation}
 expressed in the computational basis $\{ |0\rangle ,|1\rangle \}$. 

Suppose we have two qubits, which transform according to the representation $\Gamma_3\otimes \Gamma_3$.  We find that
\begin{equation}
\Gamma_3\otimes \Gamma_3=\Gamma_1 \oplus \Gamma_2 \oplus \Gamma_3 ,
\label{D3rep}
\end{equation}
where $|v_{1}\rangle = (|00\rangle + |11\rangle )/\sqrt{2}$ transforms as $\Gamma_{1}$, $|v_{2}\rangle = (|01\rangle - |10\rangle )/\sqrt{2}$ transforms as $\Gamma_{2}$, and the subspace that transforms as $\Gamma_{3}$ is spanned by $|v_{3}\rangle = (|00\rangle - |11\rangle )/\sqrt{2}$ and $|v_{4}\rangle = (|01\rangle + |10\rangle )/\sqrt{2}$.  For a state $|\phi\rangle$ in the carrier space of $\Gamma_1\oplus\Gamma_2\oplus\Gamma_3$, we have
\begin{equation}
|\phi\rangle = \sum_{j=1}^{4} c_{j}|v_{j}\rangle .
\end{equation}
We can write $|\phi\rangle$ in the form,
\begin{equation}
|\phi\rangle=c_1|u_1\rangle+c_2|u_2\rangle+\sqrt{|c_3|^2+|c_4|^2}|u_3\rangle ,
\end{equation}
where
\begin{equation}
|u_p\rangle:=|v_p\rangle,\ p=1,2;\quad 
|u_3\rangle:=\frac{c_3 |v_3\rangle+c_4|v_4\rangle}{\sqrt{|c_3|^2+|c_4|^2}},
\end{equation}
so $\|\phi_p\|=|c_p|$, for $p=1,2$, and $\|\phi_3\|=\sqrt{|c_3|^2+|c_4|^2}$.
Using Eq.~(\ref{Ps}) we find
\begin{equation}
P_{\rm s}(\rho) = \frac{1}{6} \left( |c_{1}|+|c_{2}|+\sqrt{2}\,\sqrt{|c_{3}|^{2}+|c_{4}|^{2}}\,\right)^{2} .
\end{equation}
%
%
In this case, the maximum value of $P_{\rm s}$ is $2/3$.

If we apply our duality relation to the representation in Eq.\ (\ref{D3rep}), we find $D=2/9$ and $M=\tilde M=1$ (for $\{|\eta_p\rangle\}_{p=1}^3$ orthogonal), giving us
\begin{equation}
\left( P_{{\mathcal U}_g}-\frac{2}{9}\right)^{2}  +  \left(P_{{\mathcal H}_p}-\frac{1}{3}\right)^{2} \leq  \frac{4}{9} .
\end{equation}
Note that in the case that the states $\{|\eta_p\rangle\}_{p=1}^3$ are orthogonal, we have $P_{{\mathcal U}_g}^{\rm orth}=D =2/9$ and $P_{{\mathcal H}_p}=1$, so that in this case the inequality becomes an equality.  Unlike in the case of a cyclic group, where all of the invariant subspaces are one-dimensional, and we can do no better than guessing which ${\mathcal U}_g$ we have when the states~$|\eta_{p}\rangle$ are orthogonal, in this case, since one of the subspaces is two-dimensional, there is some information about the transformation that survives ($D=2/9>1/6=1/|G|$).

\section{Conclusion}
We have derived a duality relation for finite groups, which generalizes those for wave-particle duality.  One of the quantities in the duality relations, $P_{{\mathcal U}_g}$, i.e., the probability of successfully discriminating the channels ${\mathcal U_g}$, is a measure of the asymmetry of a state under the action of the group.  If the group is cyclic, which corresponds to the usual case of phase information versus path information, $P_{{\mathcal U}_g}$ reduces to the $l_{1}$ measure of coherence.  The other quantity, $P_{{\mathcal H}_p}$, reflects our ability to discriminate the tags $\{|\eta_p\rangle\}$ attached to the irreducible representations  that act on the system.  Since irreducible representations act in invariant subspaces, the tags can be interpreted as labelling these subspaces.  If we have a network that implements the group transformations, such as that in Fig.~\ref{f-1}, the tags tell us which part of the network the particle went through, and so in the case of one-dimensional irreducible representations, they simply tell us about the path the particle took.  In the usual case, if we have complete information about the path, the quantum coherence is zero and no information about the phases is left, while in the more general case considered here, since some of the subspaces have dimension greater than one, we can know which subspace the particle went through, and there will  be some information left about the the group transformation, that is, the  state of the particle will still have some asymmetry.

\section{Acknowledgements} 

EB and JC were supported by the Spanish MINECO (grants FIS2013-40627-P and FIS2016-86681-P), with the support of FEDER funds, and by the Generalitat de Catalunya, CIRIT (projects 2014-SGR-966  and 2017-SGR-1127). EB thanks the Department of Physics and Astronomy at Hunter College for its hospitality. JB was partially supported by a PSC-CUNY Grant.

\appendix

\section{}
\label{App A}

First, we will show that $P_{\rm s}(\rho )$ decreases under G-covariant quantum maps, that is, $P_{\rm s}(\mathcal{E}(\rho )) \leq P_{\rm s}(\rho )$, where $\mathcal{E}$ is a trace-preserving, completely positive and G-covariant map.  Let the Kraus operators for $\mathcal{E}$ be $A_{j}$,
\begin{equation}
\mathcal{E}(\rho ) =\sum_{j} A_{j}\rho A^{\dagger}_{j} ,
\end{equation}
and set $\rho_{g}=U(g)\rho U^{\dagger}(g)$.  Let $\{\Pi_{g}\}_{g\in G}$ be the optimal minimum-error POVM that discriminates among the states $\rho_{g}$.  We then have that
\begin{equation}
P_{\rm s}(\rho ) = \frac{1}{|G|}\sum_{g\in G} {\rm Tr}(\rho_{g}\Pi_{g}) ,
\end{equation}
where $|G|$ is the number of elements in $G$.  Now let $\{\Pi_{g}^{(\mathcal{E})}\}_{g\in G}$ be the optimal minimum-error POVM that discriminates among the states $\mathcal{E}(\rho_{g})$.  We then have that
\begin{eqnarray}
P_{\rm s}(\mathcal{E}(\rho ) )& = & \frac{1}{|G|}\sum_{g\in G} {\rm Tr}(U(g)\mathcal{E}(\rho )U^{\dagger}(g)\Pi_{g}^{(\mathcal{E})} ) \nonumber \\
 & = &  \frac{1}{|G|}\sum_{g\in G} {\rm Tr}(\mathcal{E}(\rho_{g}) \Pi_{g}^{(\mathcal{E})}) \nonumber \\
 & = & \frac{1}{|G|}\sum_{g\in G} \sum_{j} {\rm Tr}( \Pi_{g}^{(\mathcal{E})} A_{j}\rho_{g}A_{j}^{\dagger})
 \nonumber \\
 & = & \frac{1}{|G|}\sum_{g\in G} {\rm Tr}\left( \left[ \sum_{j} A_{j}^{\dagger} \Pi_{g}^{(\mathcal{E})} A_{j}\right] \rho_{g}\right) .
\end{eqnarray}
Define 
\begin{equation}
\Pi_{g}^{\prime} = \sum_{j} A_{j}^{\dagger} \Pi_{g}^{(\mathcal{E})} A_{j} .
\end{equation}
These operators are positive and sum to the identity, and, therefore, constitute a POVM.  Because $\{\Pi_{g}\}_{g\in G}$ is the optimal minimum-error POVM that discriminates among the states $\rho_{g}$, we have that
\begin{equation}
\sum_{g\in G} {\rm Tr}(\Pi_{g}^{\prime}\rho_{g}) \leq \sum_{g\in G} {\rm Tr}(\Pi_{g}\rho_{g}) ,
\end{equation}
which implies that $P_{\rm s}(\mathcal{E}(\rho )) \leq P_{\rm s}(\rho )$ .

Next, we would like to show that $P_{\rm s}(\rho )$ is convex.  Let $\rho = \sum_{n} p_{n}\rho_{n}$ and let 
$\{\Pi_{g}^{(n)}\}_{g\in G}$ be the optimal POVM for discriminating the states $U(g)\rho_{n} U(g)^{\dagger}$ for $g\in G$.  If~$\{\Pi_{g}\, | \, g\in G\}$ is the optimal POVM for discriminating the states 
$U(g)\rho U(g)^{\dagger}$ for $g\in G$, then 
\begin{eqnarray}
\frac{1}{|G|} \sum_{g\in G} {\rm Tr}(U(g)\rho_{n}U(g)^{\dagger}\Pi_{g}) \nonumber \\
 \leq  \frac{1}{|G|} \sum_{g\in G} {\rm Tr}(U(g)\rho_{n}U(g)^{\dagger}\Pi_{g}^{(n)})  ,
\end{eqnarray}
and this implies that 
\begin{equation}
P_{\rm s}(\rho ) \leq \sum_{n} p_{n} P_{\rm s} (\rho_{n}) .
\end{equation}

\section{}
\label{App B}
Suppose we have a set of states $\{ U(g)|\phi\rangle\, |\, g\in G\}$, where $G$ is a group and $U(g)$ is a unitary representation of $G$.  We will denote the identity element of the group by $e$.  Our object is to find a POVM that will optimally discriminate among these states with minimum error.  We can assume that the POVM can be expressed as $\Pi_{g}=U(g)\Pi_{e}U(g)^{\dagger}$, where $\Pi_{g}$ is the POVM element corresponding to $U(g)|\phi\rangle$ and $\Pi_{e}$ corresponds to $|\phi\rangle$ (see Appendix C).  Assuming the states are equally likely, the success probability for the measurement is
\begin{equation}
P_{s} = \frac{1}{|G|}\sum_{g\in G} \langle\phi |U(g)^{\dagger}\Pi_{g}U(g)|\phi\rangle = \langle\phi |\Pi_{e}|\phi\rangle ,
\end{equation}
where $|G|$ is the order of the group.

As in the main text, let us assume that $U(g)$ is a representation of $G$ in which each irreducible representation appears at most once and denote the p$^{\rm th}$ irreducible representation by $\Gamma_{p}(g)$.  If $\openone_{p}$ is the projector onto the invariant subspace on which the p$^{\rm th}$ irreducible representation acts, and $|X\rangle$ is a vector, then
\begin{equation}
\frac{1}{|G|} \sum_{g\in G}U(g)|X\rangle\langle X|U(g)^{\dagger} =  \sum_{p}\frac{1}{d_{p}}\| \openone_{p}X\|^{2} \openone_{p} ,
\end{equation}
where the sum is over the irreducible representations occurring in $U(g)$, and we recall that $d_{p}$ is the dimension of the p$^{\rm th}$ irreducible representation. 

Now let $|\phi_{p}\rangle = \openone_{p}|\phi\rangle$.  We then have that
\begin{eqnarray}
\langle\phi |\Pi_{e}|\phi\rangle & = & \sum_{p,q}\langle\phi_{p}|\Pi_{e}|\phi_{q}\rangle \nonumber \\
 & = & \sum_{p,q} \langle\sqrt{\Pi_{e}}\phi_{p}|\sqrt{\Pi_{e}}\phi_{q}\rangle \nonumber \\
  & \leq & \sum_{p,q} \left(\langle \phi_{p}|\Pi_{e}|\phi_{p}\rangle \langle \phi_{q}|\Pi_{e}|\phi_{q}\rangle \right)^{1/2} \nonumber \\
  & \leq & \left( \sum_{p} \langle \phi_{p}|\Pi_{e}|\phi_{p}\rangle^{1/2} \right)^{2} .
  \label{phiPiphi}
\end{eqnarray}
Now let's make use of the fact that the sum of the POVM elements is the identity.  Let $|X_{p}\rangle$ be a vector of norm one in the invariant subspace corresponding to the p$^{\rm th}$ irreducible representation.  Then
\begin{eqnarray}
\hspace{-2em}
\sum_{g\in G}\!{\rm Tr}(\Pi_{g} |X_{p}\rangle\langle X_{p}|)\! &=&
\!{\rm Tr}\left[\left(\sum_{g\in G}\Pi_g\right)|X_p\rangle\langle X_p|\right]\nonumber\\[.5em]
&=&\! 1 .
\end{eqnarray}
However, we also have
\begin{eqnarray}
\hspace{-.5em}
\sum_{g\in G}\!{\rm Tr}(\Pi_{g} |X_{p}\rangle\langle X_{p}|)
\! &=&\!\! \sum_{g\in G}\!{\rm Tr}(U(g)\Pi_{e}U(g)^{\dagger} |X_{p}\rangle\langle X_{p}|) \nonumber\\
\!&=&\!  {\rm Tr}\!\left[\! \Pi_{e}\!\! \sum_{g\in G} \!U\!(g)^{\!\dagger}|X_{p}\rangle\langle X_{p}| U\!(g)\!\right] \nonumber \\[.5em]
 & =&\!  {\rm Tr}\!\left[ \Pi_{e}\!\!  \sum_{g\in G} \!U\!(g^{-1}\!)|X_{p}\rangle\langle X_{p}| U\!(g^{-1})^{\!\dagger}\!\right] \nonumber \\[.5em]
 & =&  \frac{|G|}{d_{p}} {\rm Tr}(\openone_{p}\Pi_{e}\openone_{p}) .
\end{eqnarray}
Therefore, 
\begin{equation}
{\rm Tr}(\openone_{p}\Pi_{e}\openone_{p}) = \frac{d_{p}}{|G|}\ .
\end{equation}
We also have that
\begin{equation}
\frac{1}{\|\phi_{p}\|^{2}} \langle \phi_{p}| \Pi_{e}|\phi_{p}\rangle \leq {\rm Tr}(\openone_{p}\Pi_{e}\openone_{p}) ,
\end{equation}
which, finally gives us that
\begin{equation}
\label{upbound}
\langle\phi |\Pi_{e}|\phi\rangle \leq \left( \sum_{p} \sqrt{\frac{d_{p}}{|G|}} \|\phi_{p}\| \right)^{2} .
\end{equation}

Now let us find a POVM that achieves this bound.  Choose $\Pi_{e}=|X\rangle\langle X|$ for some vector $|X\rangle$.  Then the requirement that the POVM elements sum to the identity gives us
\begin{equation}
\sum_{g\in G}U(g)|X\rangle\langle X|U(g)^{\dagger}= \sum_{p}\frac{|G|}{d_{p}} \|X_{p}\|^{2} \openone_{p} = \openone,
\end{equation}
where $|X_{p}\rangle = \openone_{p}|X\rangle$.  This implies that
\begin{equation}
\|X_{p}\| = \sqrt{\frac{d_{p}}{|G|}} .
\end{equation}
Now assume that we choose $|X_{p}\rangle$ parallel to $|\phi_{p}\rangle$.  This implies that
\begin{equation}
\langle\phi_{p}|X_{p}\rangle = \sqrt{\frac{d_{p}}{|G|}} \|\phi_{p}\| ,
\end{equation}
and
\begin{eqnarray}
P_{\rm s} & = & \langle\phi |\Pi_{e}|\phi\rangle = \sum_{p,q}\langle\phi |X_{p}\rangle\langle X_{q}|\phi\rangle \nonumber \\
 & = & \left( \sum_{p} \sqrt{\frac{d_{p}}{|G|}} \|\phi_{p}\| \right)^{2} .
 \label{Ps no-rep}
\end{eqnarray}
Therefore, this POVM achieves the upper bound in Eq.~(\ref{upbound}) and is the minimum-error POVM of covariant form, which is the optimal minimum-error POVM.

\section{}\label{App C}
Suppose we have a set of states $\{ U(g)|\phi\rangle\}_{g\in G}$, where~$G$ is a group and $U(g)$ is a unitary representation of $G$.  
Our object is to show that in this case we can assume with no loss of generality that the optimal POVM for discriminating the given states is of the form $\Pi_g=U(g)\Pi_eU^\dagger(g)$, which we call covariant. We will do it by proving that for any POVM, $\{\tilde\Pi_g\}_{g\in G}$, we can always find a covariant POVM that attains the very same success probability~$\tilde P_{\rm s}$.
%
Assuming that the states are equally likely we have
\begin{eqnarray}
\tilde P_{\rm s}&=&\frac{1}{|G|}\sum_{g\in G} {\rm Tr}\!\left[U(g)|\phi\rangle\langle\phi|U^\dagger(g) \, \tilde\Pi_g\right] \nonumber \\
&=&{\rm Tr}\left[ |\phi\rangle\langle\phi|\; \frac{1}{|G|}\sum_{g\in G} U^\dagger(g)\tilde\Pi_g U(g)\right] \nonumber \\
&=&{\rm Tr}\left( |\phi\rangle\langle\phi|\Omega\right)=\langle \phi|\Omega|\phi\rangle,
\label{Ptilde}
\end{eqnarray}
where we recall that  $|G|$ is the order of the group and we have defined 
\begin{equation}
\Omega=\frac{1}{|G|}\sum_{g\in G} U^\dagger(g)\tilde\Pi_g U(g) .
\end{equation}
We further define $\Pi_g=U(g)\Omega U^\dagger(g)$. Each $\Pi_g$ is positive and
\begin{eqnarray}
\sum_{g\in G}\Pi_g\!\!&=&\!\frac{1}{|G|}\!\!\sum_{g,g'\!\in G} \!\!U(g)U^\dagger(g')\tilde\Pi_{g'}U(g')U^\dagger(g)\nonumber\\
&=&\!\frac{1}{|G|}\!\!\sum_{g,g'\!\in G}\!\!U(gg'^{-1})\tilde\Pi_{g'}U^\dagger(gg'^{-1})\nonumber\\
&=&\!\frac{1}{|G|}\!\!\sum_{g''\!\!,g'\in G}\!\!\! U(g'')\tilde\Pi_{g'}U^\dagger(g'')\nonumber\\
&=&\!\frac{1}{|G|}\!\!\sum_{g''\!\in G}\!\!U(g'')\!\Bigg(\sum_{g'\in G}\tilde\Pi_{g'}\!\Bigg)U^\dagger(g'')\!=\!
\openone.
\end{eqnarray}
This shows that the set $\{\Pi_g\}_{g\in G}$ defines a proper POVM, where~$\Pi_{e}=\Omega$ is the POVM element corresponding to~$|\phi\rangle$ and $\Pi_g$  corresponds to $U(g)|\phi\rangle$. Moreover, this POVM gives the same success probability as $\tilde\Pi_g$, since
\begin{eqnarray}
P_{\rm s}&=&\frac{1}{|G|}\sum_{g\in G}{\rm Tr}\left[U(g)|\phi\rangle\langle\phi|U^\dagger(g)\Pi_g\right]\nonumber\\
&=&
\frac{1}{|G|}\sum_{g\in G}{\rm Tr}\left[|\phi\rangle\langle\phi|U^\dagger(g)\Pi_g U(g)\right]\nonumber\\
&=&
\frac{1}{|G|}\sum_{g\in G}{\rm Tr}\left(|\phi\rangle\langle\phi|\Omega\right)
=\langle\phi|\Omega|\phi\rangle=\tilde P_{\rm s},
\end{eqnarray}
as Eq.~(\ref{Ptilde}) shows. The proof also works for mixed states.


\end{document}